**PAPER**

# Electronic structure and superconductivity of the non-centrosymmetric $Sn_4As_3$

C A Marques[1], M J Neat[1], C M Yim[1], M D Watson[1], L C Rhodes[1], C Heil[2], K S Pervakov[3], V A Vlasenko[3], V M Pudalov[3,4], A V Muratov[3], T K Kim[5] and P Wahl[1]

[1] School of Physics and Astronomy, University of St Andrews, North Haugh, St Andrews, Fife KY16 9SS, United Kingdom
[2] Institute of Theoretical and Computational Physics, Graz University of Technology, NAWI Graz, 8010 Graz, Austria
[3] Ginzburg Center for High Temperature Superconductivity and Quantum Materials, Lebedev Physical Institute, Moscow, 119991, Russia
[4] National Research University Higher School of Economics, Moscow 101000, Russia
[5] Diamond Light Source, Harwell Campus, Didcot OX11 0DE, United Kingdom

E-mail: wahl@st-andrews.ac.uk

## Abstract

In a superconductor that lacks inversion symmetry, the spatial part of the Cooper pair wave function has a reduced symmetry, allowing for the mixing of spin-singlet and spin-triplet Cooper pairing channels and thus providing a pathway to a non-trivial superconducting state. Materials with a non-centrosymmetric crystal structure and with strong spin–orbit coupling are a platform to realize these possibilities. Here, we report the synthesis and characterisation of high quality crystals of $Sn_4As_3$, with non-centrosymmetric unit cell ($R3m$). We have characterised the normal and superconducting states using a range of methods. Angle-resolved photoemission spectroscopy shows a multiband Fermi surface and the presence of two surface states, confirmed by density-functional theory calculations. Specific heat measurements reveal a superconducting critical temperature of $T_c \sim 1.14$ K and an upper critical magnetic field of $\mu_0 H_c \gtrsim 7$ mT, which are both confirmed by ultra-low temperature scanning tunneling microscopy and spectroscopy. Scanning tunneling spectroscopy shows a fully formed superconducting gap, consistent with conventional s-wave superconductivity.

## 1. Introduction

Identification of a spin-triplet superconductor would provide us with a potential solid state platform for topological quantum computations, a variant that is particularly robust against decoherence—one of the main impediments to realization of larger scale quantum calculations. There are different routes to realizing spin-triplet superconductivity (SC): either through engineered heterostructures, or in materials where triplet pairing is allowed or even promoted. Here, we focus on the latter path. One class of materials where a triplet component becomes allowed are non-centrosymmetric SC, where mixing of spin-singlet and spin-triplet Cooper pairing channels is possible and Rashba spin–orbit coupling (SOC) can lead to a lifting of Kramers degeneracy for electronic states in the bulk of the material [1]. A possible triplet component is expected to manifest in a number of observables: the upper critical magnetic field will be much higher than for a singlet superconductor and the SC gap will exhibit a more complex structure than the hard gap predicted by the Bardeen–Cooper–Schrieffer (BCS) theory for a singlet SC. One would also expect topologically protected bound states near defects and boundaries that could be detected in local measurements. Experimentally, evidence for this mixing has been found in the non-centrosymmetric heavy fermion superconductor CePt$_3$Si, where the strong SOC gives rise to an SC gap with line nodes [2]. However, the degree of mixing is not only determined by the strength of the SOC but also by the dominant pairing interaction [3]. If spin-singlet pairing interactions are dominant, the SC in the non-centrosymmetric material will follow the predictions by the BCS theory, as has been found in the case of BiPd [4], and it is independent of the SOC strength. Nevertheless, in BiPd the breaking of inversion symmetry together with strong SOC leads to Dirac-cone surface states [5] with an intricate spin texture [6].

In Sn-based compounds, the development of unconventional SC has been suggested, particularly in the case of the topological crystalline insulator SnTe with indium doping, where strong SOC has been shown to play an important role [7, 8]. A non-centrosymmetric crystal structure in Sn-based materials could then, in principle, favour the appearance of a triplet component. Here we focus on the non-centrosymmetric material Sn$_4$As$_3$. Transport measurements reveal a metallic nature [9] and SC critical temperature, $T_c$, in the range 1.16–1.19 K [10].

It was only recently that the crystal structure of Sn$_4$As$_3$ was identified as belonging to the non-centrosymmetric space group $R3m$ (no. 160), with a hexagonal unit cell [11]. Additionally, band structure calculations show several bands crossing the Fermi level, with a depletion of the density of states (DOS) at $\sim$0.4 eV above the Fermi energy [11]. However, a direct measure of its electronic structure has been missing. Moreover, there have been no recent reports on its SC, apart from other SnAs-based superconductors such as SnAs [12, 13] and NaSn$_2$As$_2$ [14–16] which exhibit centrosymmetric crystal structures, revealing SC consistent with spin-singlet pairing.

In this work, we report a detailed study of the properties of single crystal samples of Sn$_4$As$_3$, through thermodynamic and spectroscopic characterisation of both normal and SC states using angle-resolved photoemission spectroscopy (ARPES) and ultra-low temperature scanning tunneling microscopy and spectroscopy (STM/STS). The experimental results from ARPES and STM are directly compared with bulk and slab density functional theory (DFT) calculations.

## 2. Methods

*Sample Growth.* Sn$_4$As$_3$ single crystals were synthesized from high purity elemental Sn (99.99%) and As (99.9999%), weighted in stoichiometric molar ratio (4:3). Synthesis was performed inside a quartz ampoule with Ar atmosphere at a residual pressure of 0.2 bar. The ampoule was put into a furnace and heated up to 600 °C for 24 h. The temperature was then increased to 650 °C, followed by slow cooling down to room temperature at a rate of 2 °C h$^{-1}$. The samples were characterised by energy-dispersive x-ray spectrometry (EDS) and x-ray diffraction (XRD), revealing a chemical composition of Sn$_{3.8}$As$_3$ and lattice constants of $a = 4.0891$ Å and $c = 36.0524$ Å, in agreement with references [9, 11].

*Scanning tuneling microscopy and spectroscopy.* STM measurements were performed with a home-built ultra-low temperature STM, mounted in a dilution refrigerator [17] with a base temperature of 10 mK and in a superconducting magnet with maximum field of 14 T. The Pt–Ir tip was cut from a wire and conditioned by field-emission on an Au single crystal prior to measuring. The Sn$_4$As$_3$ sample was cleaved *in situ* at low temperatures ($T \approx 20$ K). The bias voltage was applied to the sample. Differential conductance spectra were recorded using a lock-in amplifier ($f = 437$ Hz), with amplitudes of modulations set at 15 mV and 25 μV for measurements at 11 K and 50–900 mK, respectively.

*Angle-resolved photoemission spectroscopy.* ARPES measurements were performed at the I05 beamline of Diamond Light Source, UK [18]. Single-crystal samples were cleaved *in situ* in a vacuum better than $2 \times 10^{-10}$ mbar and measured at temperatures of 20 K. Measurements were performed using linear horizontal (LH) and linear vertical (LV) polarized synchrotron light with variable photon energy, using a Scienta R4000 hemispherical electron energy analyzer with an angular resolution of 0.2° and an energy resolution of 20 meV.

*Density functional theory calculations.* Bulk electronic band structure calculations were performed for the experimental crystal structure of Sn$_4$As$_3$ from Kovnir *et al* [11] in the generalized gradient approximation (GGA) using WIEN2k [19], taking into account SOC. These were used to produce a three dimensional (3D) Fermi surface, as well as 2D cuts at different $k_z$ planes. Additionally, bulk and slab calculations were carried out with the Quantum Espresso package [20] using GGA within the framework of Perdew–Burke–Ernzerhof [21], employing optimized norm-conserving Vanderbilt pseudopotentials [22, 23], for the same crystal structure as before. We chose a plane wave (PW) cutoff of 80 Ry, Gaussian smearings of 0.02 Ry, and a $24 \times 24 \times 6$ Monkhorst–Pack $k$-grid to sample the Brillouin zone (BZ). Both DFT codes yielded consistent results for the bulk electronic structure. To simulate STM images of the pristine surface and of the surface with a Sn vacancy, we considered a $3 \times 3 \times 1$ supercell, while to simulate

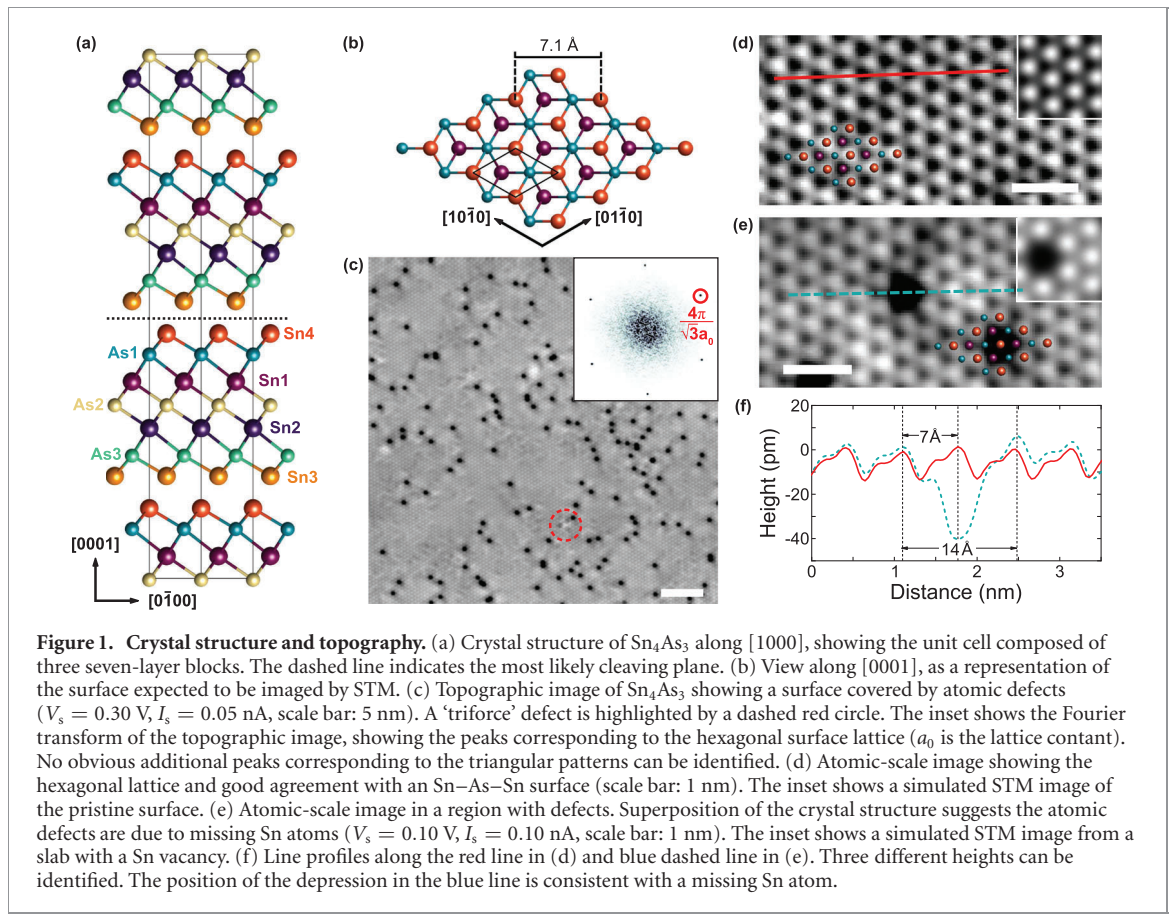

**Figure 1. Crystal structure and topography.** (a) Crystal structure of $Sn_4As_3$ along [1000], showing the unit cell composed of three seven-layer blocks. The dashed line indicates the most likely cleaving plane. (b) View along [0001], as a representation of the surface expected to be imaged by STM. (c) Topographic image of $Sn_4As_3$ showing a surface covered by atomic defects ($V_s = 0.30$ V, $I_s = 0.05$ nA, scale bar: 5 nm). A 'triforce' defect is highlighted by a dashed red circle. The inset shows the Fourier transform of the topographic image, showing the peaks corresponding to the hexagonal surface lattice ($a_0$ is the lattice contant). No obvious additional peaks corresponding to the triangular patterns can be identified. (d) Atomic-scale image showing the hexagonal lattice and good agreement with an Sn–As–Sn surface (scale bar: 1 nm). The inset shows a simulated STM image of the pristine surface. (e) Atomic-scale image in a region with defects. Superposition of the crystal structure suggests the atomic defects are due to missing Sn atoms ($V_s = 0.10$ V, $I_s = 0.10$ nA, scale bar: 1 nm). The inset shows a simulated STM image from a slab with a Sn vacancy. (f) Line profiles along the red line in (d) and blue dashed line in (e). Three different heights can be identified. The position of the depression in the blue line is consistent with a missing Sn atom.

STM images with Sn or As vacancies at different subsurface layers, we also considered larger $4 \times 4 \times 1$ supercells. The BZs were sampled using Monkhorst–Pack $k$-grids ($5 \times 5 \times 1$ for the $3 \times 3 \times 1$ supercell, $4 \times 4 \times 1$ for the $4 \times 4 \times 1$ supercell). We chose a vacuum region of 10 Å in the case of the slab calculations, SOC was neglected for all STM simulations and all DOS calculations have been performed with a denser $36 \times 36 \times 8$ ($36 \times 36 \times 1$ for the slab) BZ grid and a Gaussian smearing of 0.01 Ry.

*Specific heat measurements.* Specific heat of $Sn_4As_3$ crystals was measured by thermal relaxation technique, using a PPMS-9 (Quantum Design) with a $^3$He calorimeter. The mass of the sample was $m = 11$ mg. Measurements were performed at temperatures 0.37–2 K and magnetic fields of 0–20 mT.

## 3. Results

### 3.1. Crystal structure and surface topography

The crystal structure of $Sn_4As_3$ belongs to the non-centrosymmetric group $R3m$, whose hexagonal unit cell is shown in figure 1(a). The unit cell is composed of three seven-layer blocks of alternating Sn and As layers stacked along the $c$-axis. Inside each block, pairs of atoms that would otherwise be symmetrically equivalent (Sn1 and Sn2; Sn3 and Sn4; As1 and As3) are inequivalent: the Sn atoms in different layers form distorted octahedra with the surrounding atoms, which are responsible for the lack of inversion centre [11]. The weakest bond in the unit cell occurs between Sn3–Sn4 atoms from two different blocks (indicated by the dashed line in figure 1(a)), with the larger bond distance of 3.24 Å, comparable to that observed in other layered SnAs-based materials [14]. Thus, the crystal is expected to cleave well in the (0001) plane, between the Sn3–Sn4 layers. The exposed surface can be either Sn4 or Sn3, which are structurally identical. A Sn4-terminated surface is illustrated in figure 1(b).

A typical STM topographic image with atomic resolution is shown in figure 1(c). Besides an atomic lattice with hexagonal symmetry, atomic defects can be identified, which cover roughly 1.76% of the surface. Less abundant defects shaped as a 'triforce' triangle (indicated by a dashed red circle) can also be observed, which usually appear once in every (30 nm)$^2$ field of view. The Fourier transform (FT) reveals the expected Bragg peaks for a hexagonal lattice (inset of figure 1(c)). A high-resolution topography is presented in figure 1(d), where three different levels of contrast can be distinguished. Simulated STM images from DFT slab calculations confirm that the bright spots are Sn atoms and the grey spots are As atoms (see inset of figure 1(d)). In figure 1(e), a topography showing an atomic defect in detail is presented, where it can be seen that it sits at the Sn lattice site. These defects are consistent with a Sn vacancy in the topmost Sn layer, as confirmed by DFT slab calculations (inset figure 1(e)). This is also evidenced by the line profiles shown in figure 1(f), which altogether confirm that the depression in figure 1(e) is centered at a Sn4 site, hence corresponding to a Sn vacancy.

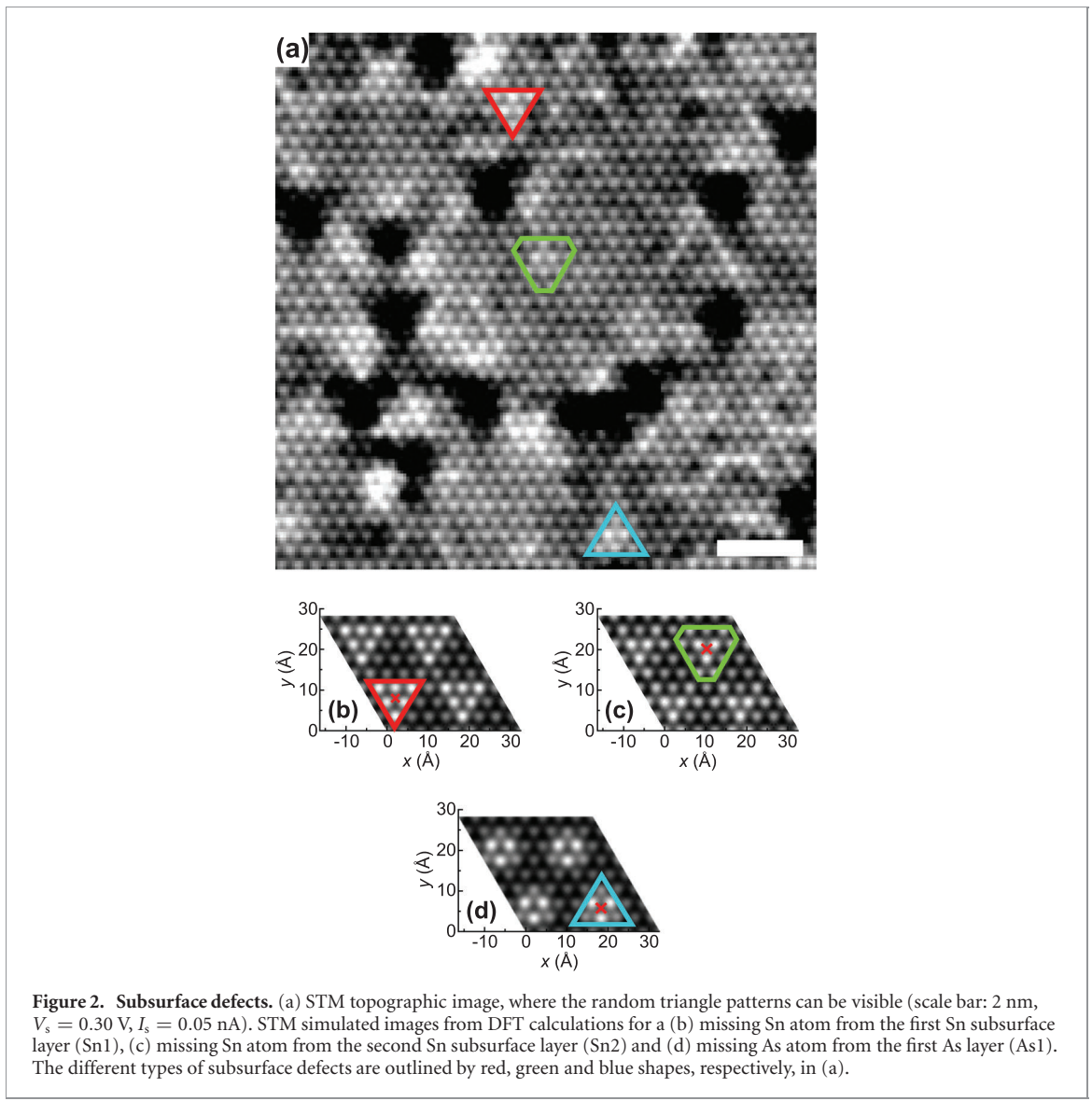

**Figure 2. Subsurface defects.** (a) STM topographic image, where the random triangle patterns can be visible (scale bar: 2 nm, $V_s = 0.30$ V, $I_s = 0.05$ nA). STM simulated images from DFT calculations for a (b) missing Sn atom from the first Sn subsurface layer (Sn1), (c) missing Sn atom from the second Sn subsurface layer (Sn2) and (d) missing As atom from the first As layer (As1). The different types of subsurface defects are outlined by red, green and blue shapes, respectively, in (a).

Additionally, triangular patterns with different intensities which are superimposed on the periodic atomic lattice can be identified, figure 2(a). FT analysis of large scale images does not show any distinct peaks other than the atomic peaks (inset of figure 1(c)), suggesting that these patterns do not have a well-defined periodicity. DFT slab calculations for different subsurface defects are shown in figure 2(b) for a missing Sn atom from the Sn1 layer, figure 2(c) from the Sn2 layer and figure 2(d) for a missing As atom from the As1 layer. These subsurface defects generate bright triangular shaped patterns at the topmost Sn layer, which could give rise to the observed patterns when these defects are randomly distributed throughout the crystal.

### 3.2. Fermi surface and electronic band structure

ARPES measurements in figures 3(a) and (b) reveal a complex Fermi surface with strong 3D character, despite the layered structure of the material. Figure 3(a) shows a Fermi surface map measured with a photon energy of $h\nu = 98$ eV. Three distinct shapes can be identified, which we label $\delta$ (green), $\beta$ (red) and $\gamma$ (blue). Both $\beta$ and $\gamma$ bands are three-fold symmetric, reflecting the trigonal symmetry of the crystal. Additionally, features with lower intensity can be observed, which are rotated by 60°. These correspond to a different $k_z$-plane of the Fermi surface, whose contribution can be increased by measuring with a photon energy of $h\nu = 120$ eV, shown in figure 3(b). Here, we see the β and γ bands appearing with six-fold symmetry, where contributions from different $k_z$-planes are marked with solid and dashed lines.

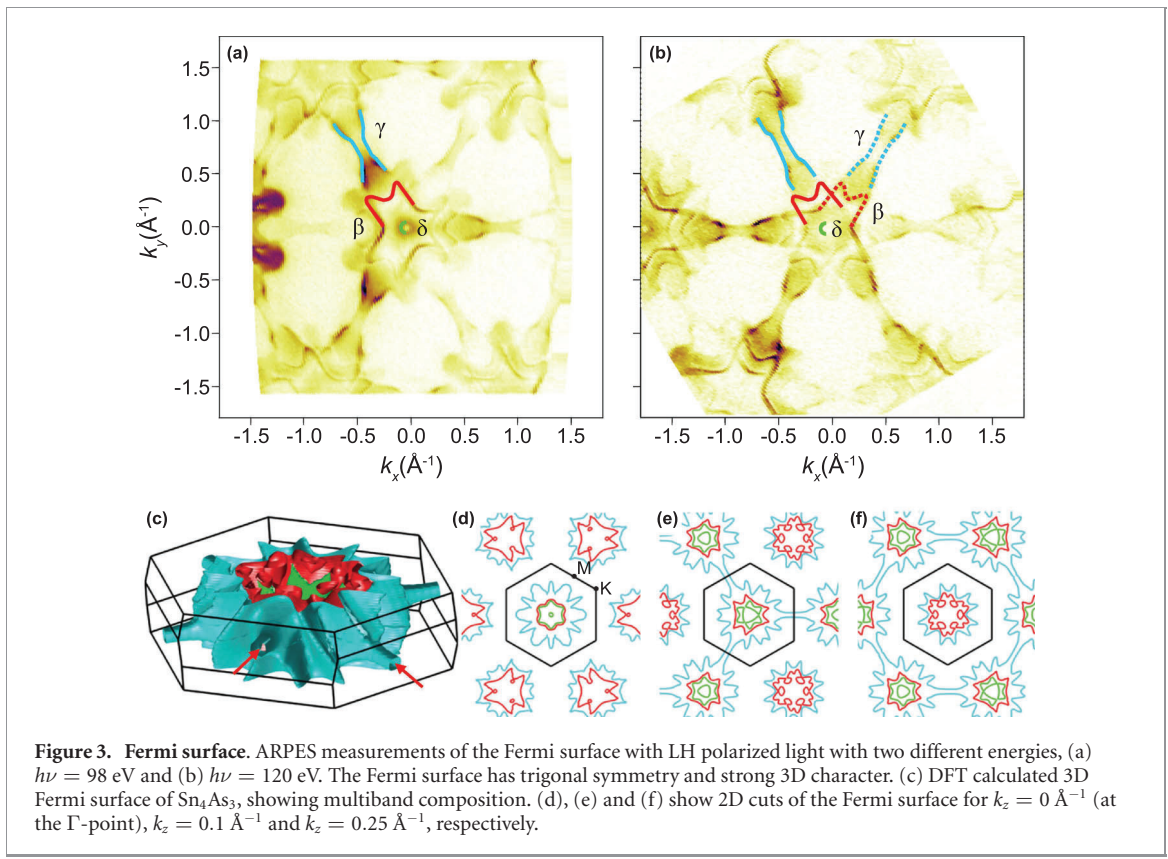

**Figure 3. Fermi surface**. ARPES measurements of the Fermi surface with LH polarized light with two different energies, (a) $h\nu = 98$ eV and (b) $h\nu = 120$ eV. The Fermi surface has trigonal symmetry and strong 3D character. (c) DFT calculated 3D Fermi surface of $Sn_4As_3$, showing multiband composition. (d), (e) and (f) show 2D cuts of the Fermi surface for $k_z = 0$ Å$^{-1}$ (at the $\Gamma$-point), $k_z = 0.1$ Å$^{-1}$ and $k_z = 0.25$ Å$^{-1}$, respectively.

Figure 3(c) shows the calculated 3D Fermi surface composed of three different bands, mainly of Sn-p and As-p character, which we can identify with the same color code and notation as above. The 3D character can be clearly seen where the tube-like features of the γ band at $k_z = 0.1$ Å$^{-1}$ above the Γ-point are 60° rotated in relation to the corresponding features at $k_z = -0.1$ Å$^{-1}$ (red arrows), reflecting the three-fold symmetry. Figures 3(d), (e) and (f) show cuts of the Fermi surface at increasing $k_z$, starting from the Γ-point. Here we project the 3D BZ into the hexagonal 2D BZ, with the high-symmetry points *M* and *K*. At the centre of the BZ the Fermi surface has six-fold symmetry, but as we move away from the centre, the three-fold symmetry becomes evident. Comparing the calculated 2D cuts with the ARPES Fermi surfaces in figures 3(a) and (b), similar features can be identified, showing good agreement between theory and experiment.

The ARPES electronic band dispersions along the $\overline{M}-\overline{\Gamma}$ and $\overline{\Gamma}-\overline{K}$ directions are shown in figure 4(a). The measurements are consistent with $Sn_4As_3$ being metallic, with several bands crossing the Fermi level, in agreement with Kovnir *et al* [11]. DFT calculations for the bulk band structure are shown in figure 4(b). Including SOC (dashed blue lines) changes only slightly the energies of the bands, mainly around the Γ-point. Overall, there is good agreement between experiment and calculations over an energy range of several electronvolts, indicating a weakly correlated nature of the $Sn_4As_3$ electronic structure. Small discrepancies between experimental data and calculations arise from the strong 3D character of the electronic dispersion and finite $k_z$-averaging in the photoemission experiment. In addition to the bulk bands predicted by the DFT calculations, the measurements show additional bands at energies close to $-1$ eV (indicated by a white arrow), which are split by $\sim 100$ meV. These two bands appear clearly in the slab DFT calculations without SOC (red lines in figure 4(c)), similar to those for SnAs [13]. The consistency with ARPES data confirms that they are two surface states (SS) separated in energy by $\sim 108$ meV. Slab calculations including SOC (dashed blue lines in figure 4(c)) reveal that each one of these SS is in addition spin-split by up to 50 meV.

Figure 4(d) shows a comparison between the total DOS from the bulk and slab DFT calculations (including SOC), the local DOS from DFT at 10 Å above the surface, the spectral function from ARPES integrated over both symmetry directions and an STM tunneling spectrum. The integrated DOS extracted from the ARPES data shows good overall agreement with the bulk total DOS. Both show finite DOS at the Fermi energy. Differences are likely because the DOS obtained from DFT calculations is from averaging over the complete BZ, whereas the one extracted from ARPES is only from integration along two high-symmetry directions.

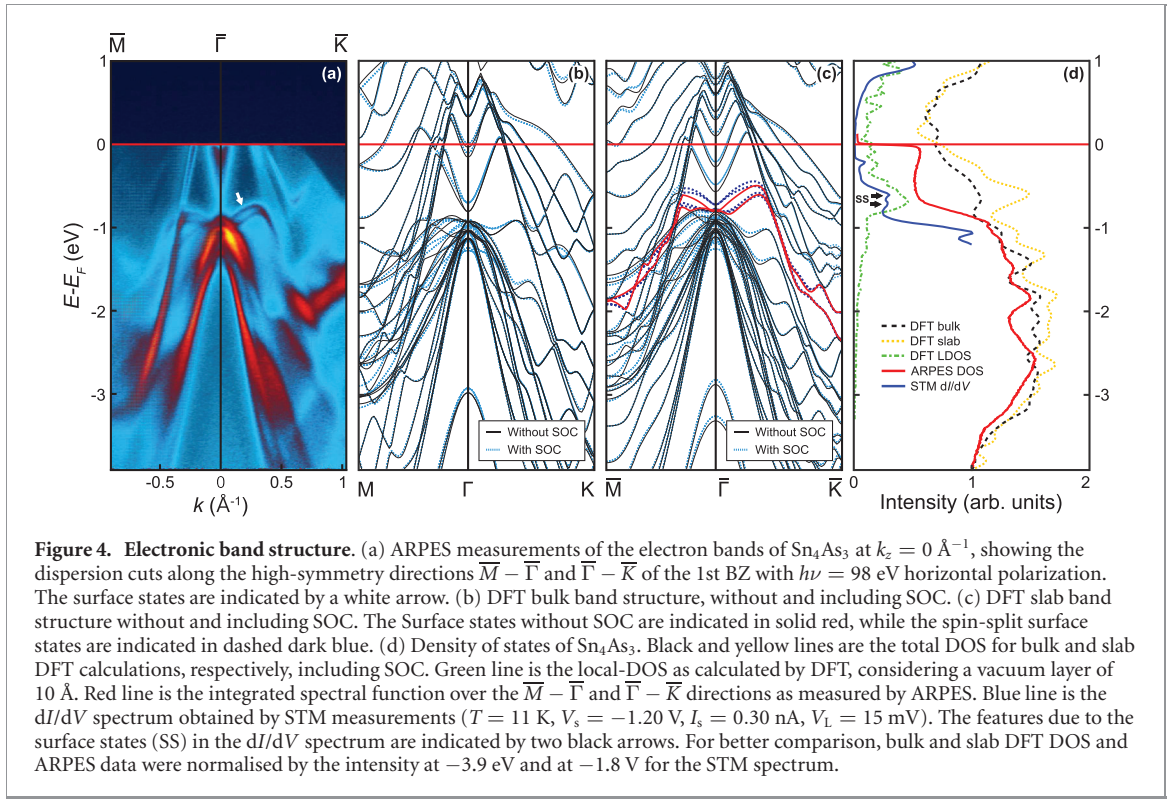

**Figure 4.** **Electronic band structure**. (a) ARPES measurements of the electron bands of $Sn_4As_3$ at $k_z = 0$ Å$^{-1}$, showing the dispersion cuts along the high-symmetry directions $\overline{M} - \overline{\Gamma}$ and $\overline{\Gamma} - \overline{K}$ of the 1st BZ with $h\nu = 98$ eV horizontal polarization. The surface states are indicated by a white arrow. (b) DFT bulk band structure, without and including SOC. (c) DFT slab band structure without and including SOC. The Surface states without SOC are indicated in solid red, while the spin-split surface states are indicated in dashed dark blue. (d) Density of states of $Sn_4As_3$. Black and yellow lines are the total DOS for bulk and slab DFT calculations, respectively, including SOC. Green line is the local-DOS as calculated by DFT, considering a vacuum layer of 10 Å. Red line is the integrated spectral function over the $\overline{M} - \overline{\Gamma}$ and $\overline{\Gamma} - \overline{K}$ directions as measured by ARPES. Blue line is the d$I$/d$V$ spectrum obtained by STM measurements ($T = 11$ K, $V_s = -1.20$ V, $I_s = 0.30$ nA, $V_L = 15$ mV). The features due to the surface states (SS) in the d$I$/d$V$ spectrum are indicated by two black arrows. For better comparison, bulk and slab DFT DOS and ARPES data were normalised by the intensity at $-3.9$ eV and at $-1.8$ V for the STM spectrum.

The DOS of the slab DFT calculation including SOC shows an increase in intensity at the energy of the SS. The STM differential conductance (d$I$/d$V$) spectrum can be taken as proportional to the local DOS. In the d$I$/d$V$ measurements (figure 4(d)), two peaks can be identified in the energy range corresponding to the SS, separated by 110 meV (indicated by black arrows), consistent with ARPES. This can be directly compared to the local DOS calculations, which shows the increase in intensity at the SS energies. The STM spectrum shows a depletion of the LDOS around the Fermi level, evidenced by a low differential conductance, which is also consistent with the calculations of the local DOS.

### 3.3. Superconductivity
*3.3.1. Thermodynamic measurements*

Specific heat measurements of $Sn_4As_3$ are shown in figure 5. In zero applied magnetic field (figure 5(a)), a clear jump in $C/T$ is observed at temperatures close to 1.1 K, typical of a superconducting transition. From the local entropy conservation, the critical temperature was found to be $T_c = 1.14 \pm 0.01$ K, close to the reported values of 1.16–1.19 K [10]. The small width of the superconducting transition is indicative of the high quality of the sample.

At low temperatures (well below the Debye temperature) the specific heat of a metal can be written as $C/T = \gamma_n + \beta T^2$, where $\gamma_n$ and $\beta$ are the electronic and the phonon contributions, respectively. The $C/T$ curve for the normal state at 20 mT field (which fully suppresses the superconducting transition) is shown in figure 5(a). It has a parabolic shape, consistent with a metallic behaviour. A second order polynomial fit (red solid line) yields $\gamma_n = 6.66 \pm 0.20$ mJ mol$^{-1}$K$^{-2}$ and $\beta = 0.933 \pm 0.143$ mJ mol$^{-1}$K$^{-4}$.[5]

The electronic specific heat, $C_{el}/T$, can be obtained by subtracting the phonon contribution. Figure 5(b) shows $C_{el}/\gamma_n T$ as a function of $T$, at zero magnetic field. The jump in specific heat at this temperature is $\Delta C_{el} = 9.74$ mJ mol$^{-1}$K$^{-1}$. The relative magnitude of the jump, $\Delta C_{el}/\gamma_n T_c = 1.30$, is close to the BCS prediction of $\Delta C_{el}/\gamma_n T_c = 1.43$. The red line in figure 5(b) shows the fit of the electronic specific heat in the superconducting state derived from the BCS theory:

$$\frac{C_{el}}{\gamma_n T} = \int_{-\infty}^{+\infty} \frac{\partial f(\varepsilon)}{\partial \varepsilon} \Re \frac{|\varepsilon|}{\sqrt{\varepsilon^2 - \Delta^2(T)}} \left(\frac{d\Delta^2(T)}{dT} - \frac{2}{T}\varepsilon^2\right) d\varepsilon, \quad (1)$$

with the temperature dependence of the gap described by

$$\Delta(T) = \Delta_0 \tanh\left(a\frac{\pi}{2}\sqrt{\frac{T_c}{T} - 1}\right), \quad (2)$$

where $\Delta_0 = r k_B T_c$ is the superconducting gap at $T = 0$ K. $f(\varepsilon)$ is the Fermi function and $a = 1.138$ obtained from fitting the BCS mean field behaviour [4] with equation 2. Fitting $r = \Delta_0/k_B T_c$ yields $r = 1.76 \pm 0.02$[6] in excellent agreement with BCS theory. Using the measured $T_c = 1.14$ K and the BCS approximation, the superconducting gap is $\Delta_0 = 1.76 k_B T_c = 0.177 \pm 0.003$ meV. Use of more complex models (introducing anisotropy or using two gaps) does not give significant improvement of the fits.

---

[5] Errors from 95% confidence bounds of the parabolic fit.

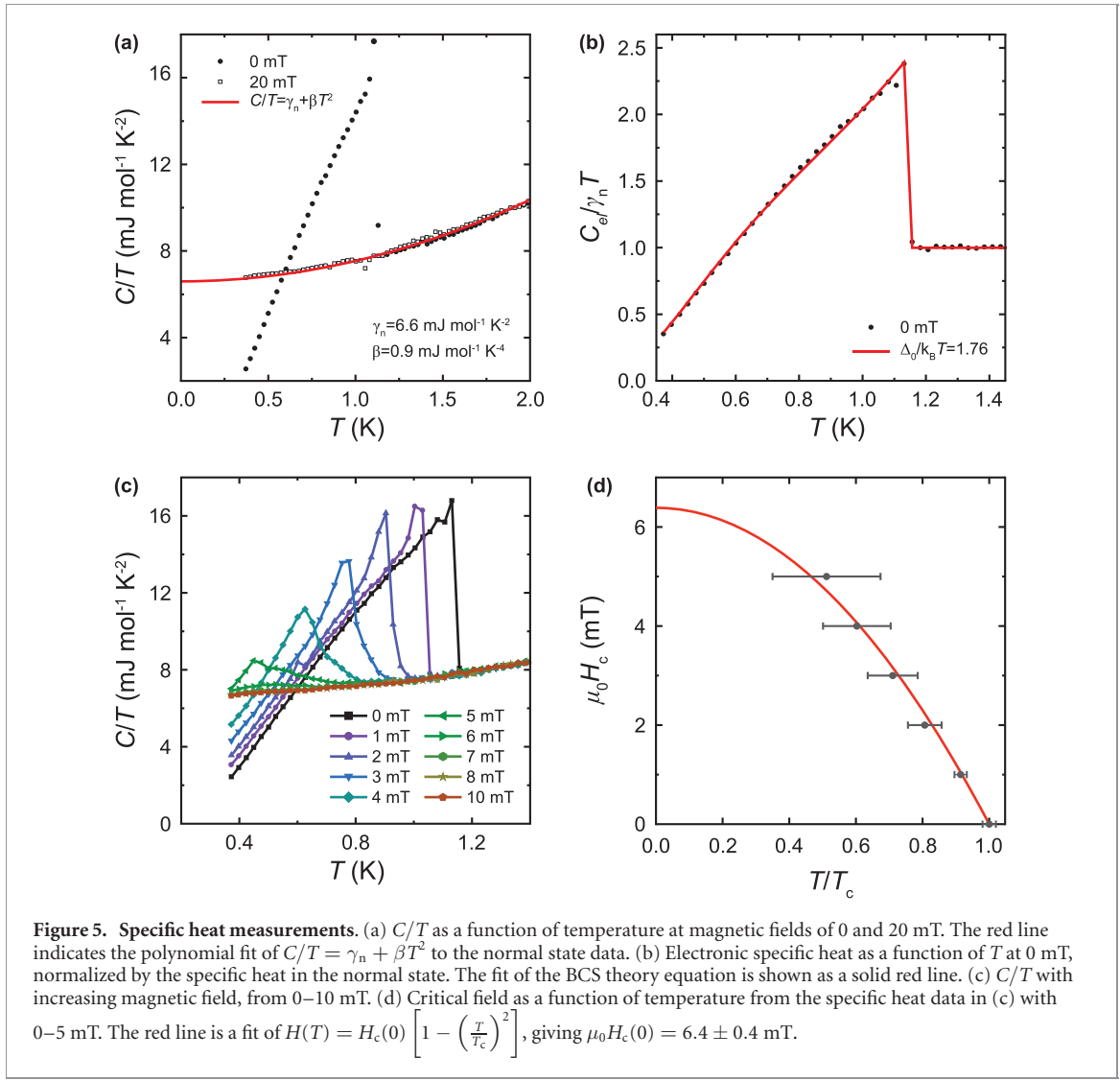

**Figure 5. Specific heat measurements**. (a) $C/T$ as a function of temperature at magnetic fields of 0 and 20 mT. The red line indicates the polynomial fit of $C/T = \gamma_n + \beta T^2$ to the normal state data. (b) Electronic specific heat as a function of $T$ at 0 mT, normalized by the specific heat in the normal state. The fit of the BCS theory equation is shown as a solid red line. (c) $C/T$ with increasing magnetic field, from 0–10 mT. (d) Critical field as a function of temperature from the specific heat data in (c) with 0–5 mT. The red line is a fit of $H(T) = H_c(0)\left[1 - \left(\frac{T}{T_c}\right)^2\right]$, giving $\mu_0 H_c(0) = 6.4 \pm 0.4$ mT.

The magnetic field dependence of $C/T$ is shown in figure 5(c). Superconductivity is found to be completely suppressed already in magnetic fields of $\mu_0 H_c \sim 7$ mT. From this data, the transition temperature can be extracted as a function of applied magnetic field, shown in figure 5(d). The upper critical field, $H_c$, can be estimated by fitting the field-dependence of the critical temperature of a type I superconductor, $H_c(T) = H_c(0)\left[1 - \left(\frac{T}{T_c}\right)^2\right]$ (red line in figure 5(d)), giving an upper critical field of $\mu_0 H_c(0) = 6.4 \pm 0.4$ mT[7].

---

[6] Error from 95% confidence bounds from the fit.

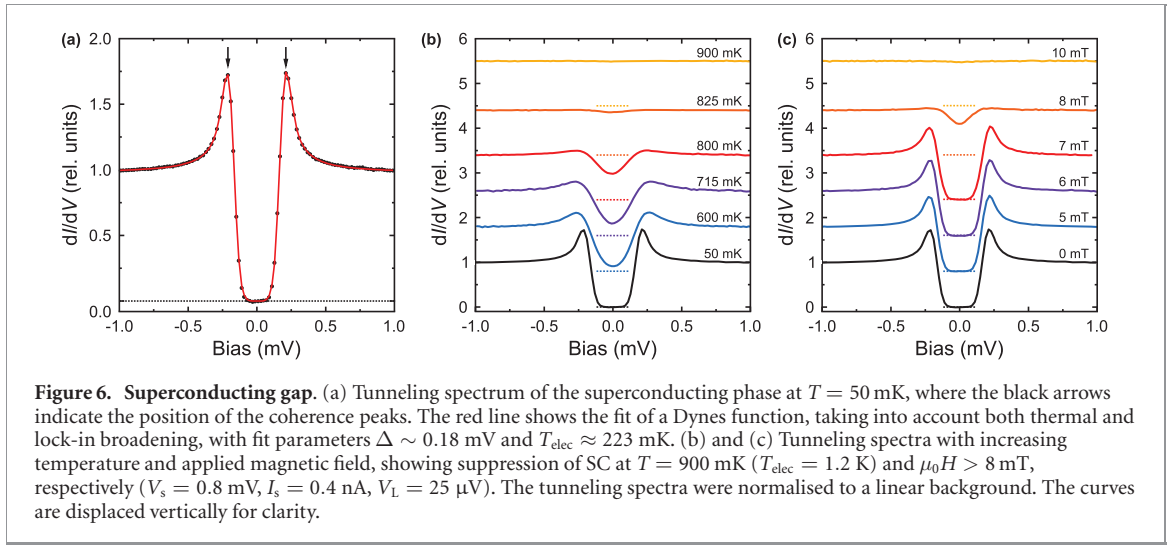

**Figure 6. Superconducting gap**. (a) Tunneling spectrum of the superconducting phase at $T = 50$ mK, where the black arrows indicate the position of the coherence peaks. The red line shows the fit of a Dynes function, taking into account both thermal and lock-in broadening, with fit parameters $\Delta \sim 0.18$ mV and $T_{\text{elec}} \approx 223$ mK. (b) and (c) Tunneling spectra with increasing temperature and applied magnetic field, showing suppression of SC at $T = 900$ mK ($T_{\text{elec}} = 1.2$ K) and $\mu_0 H > 8$ mT, respectively ($V_s = 0.8$ mV, $I_s = 0.4$ nA, $V_L = 25$ μV). The tunneling spectra were normalised to a linear background. The curves are displaced vertically for clarity.

*3.3.2. Superconducting gap*

In order to obtain further evidence of the superconducting gap structure, we have performed STM/STS measurements at temperatures below 1 K. A well resolved superconducting gap is observed in high resolution tunneling spectra, d$I$/d$V$, taken in an energy range of $\pm 1$ mV at 50 mK, shown in figure 6(a). The coherence peaks can be clearly identified, while the DOS is completely suppressed around the Fermi energy, as expected from BCS theory with $s$-wave symmetry. A Dynes equation [24] for a single isotropic gap was fitted to the data, taking into account both thermal and lock-in broadening [17]. The fitting parameters were the superconducting gap $\Delta$ and the electronic temperature, $T_{\text{elec}}$. Here, the additional broadening in the Dynes equation, $\Gamma$, was fixed to be very small ($\Gamma \sim 10^{-4}$ meV). The fit yielded $\Delta = 0.182 \pm 0.018$ meV[7] and $T_{\text{elec}} \approx 223$ mK. The electronic temperature is dominated by the lock-in modulation ($V_L = 25$ μV RMS) used in the experiment, whose contribution to broadening is larger than the thermal broadening. Using the BCS relation $\Delta/k_B T_c = 1.76$, the gap size yields a critical temperature of $T_c = 1.20 \pm 0.14$ K, which is consistent with the reported values [10] and in excellent agreement with the specific heat measurements.

Figure 6(b) shows the temperature dependence of the superconducting gap in the temperature range 50–900 mK. The d$I$/d$V$ spectra show a decrease in the gap height and movement of the coherence peaks away from the Fermi energy, consistent with thermal broadening. It can be seen that the SC is suppressed already at a temperature of 900 mK, lower than the expected temperature from the thermodynamic measurements. The apparent lower critical temperature can be due to both thermal and lock-in modulation broadening.

The magnetic field dependence of the superconducting gap at $T = 50$ mK can be seen in figure 6(c). The measurements show suppression of SC above magnetic fields of 8 mT, again consistent with the specific heat measurements.

## 4. Discussion

STM topographies show a surface consistent with a cleave between adjacent Sn layers, where the bond between atoms is expected to be weakest. The surface shows atomic defects that we identify as Sn vacancies in the topmost surface layer from comparison with DFT slab calculations and identification of the defect site. The occurrence of these defects is consistent with the chemical composition of $Sn_{3.8}As_3$ determined from *post*-growth compositional analysis by EDS.

In addition to these defects in the top surface layer, we find a random distribution of bright triangular patterns. The lack of periodicity indicates that they are not due to the presence of a charge density wave.

---

[7] Error from 95% confidence bounds from the fit.

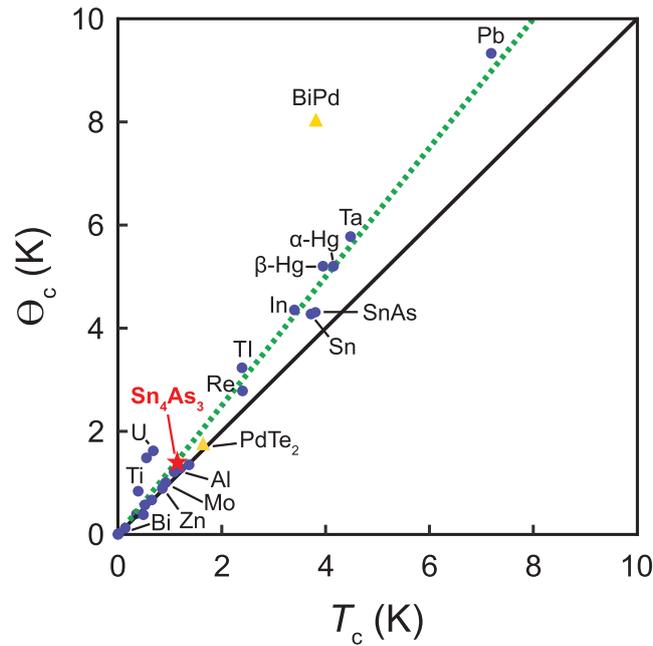

**Figure 7. Comparison to other type I superconductors.** We plot $\Theta_c = 2\mu_0 H_c(0)/\sqrt{1.43\mu_0\gamma_n}$ against $T_c$ for 22 elemental type I superconductors (blue symbols). $Sn_4As_3$ is shown as a red star. Black solid line is the expectation from BCS theory in the weak coupling limit, the green dashed line for the $\Delta/k_B T_c$ ratio of lead as an example for a strong coupling superconductor. We note that points for elemental type II superconductors such as V or Nb are far outside the range of the *y*-axis shown (at $\Theta_c = 43.3$ K and 45.6 K, respectively). For details see supplemental material. For comparison, we have included data points for BiPd and $PdTe_2$ as yellow triangles. The former has been previously determined to be a type II superconductor, while for $PdTe_2$ it has been controversial whether it is type I or type II [29, 30]. BiPd clearly violates the behaviour expected for type I superconductivity. We have also included $TaSi_2$ [31] and SnAs [13].

Additionally, bias dependent imaging and d$I$/d$V$ spectroscopy maps (not shown) reveal that these are static in energy, suggesting that they are not generated from quasiparticle scattering off defects. Simulated STM images from slab DFT calculations show that missing Sn and As atoms from deeper layers produce bright triangular shapes at the topmost layer, which resemble the observed patterns. Thus, we attribute the origin of these patterns as coming from randomly distributed defects throughout different layers of the sample. The less abundant 'triforce' defect does not seem to be captured by these calculations.

The ARPES measurements confirm the metallic nature of the material with several bands crossing the Fermi level, consistent with tunneling spectra, specific heat and DFT calculations. The Fermi surface shows significant dispersion along all directions, including *z* direction, which is evidence of a 3D character of the electronic structure despite the layered crystal structure. The ARPES measurements do reveal a pair of surface states that look at first like Rashba-spin split states, but are fully captured in calculations without SOC. Comparison with bulk and slab DFT calculations reveal that this splitting is due only to the symmetry breaking at the surface. Including SOC in the slab DFT calculations reveals that the surface states are further spin-split by up to 50 meV. While this value is within the energy resolution of our ARPES measurements, we do not observe such a splitting in experiments. One possible source for this discrepancy could be the fact that we did not perform a full geometrical relaxation for all the atoms in the slab structure.

Despite the non-centrosymmetric crystal structure of the material, the SC properties are found to be fully consistent, within the experimental errors, with what would be expected from BCS theory. The STM tunneling spectra shows a fully formed gap, which is spatially uniform and has the shape characteristic of an isotropic *s*-wave SC gap. These results follow the trend of other SnAs-based superconductors, where conventional BCS SC with *s*-wave symmetry has been found [12, 13, 15, 16].

To determine whether the material is a type-I or type-II superconductor, we compare the specific heat jump to the slope of the critical field near $T_c$. For a type-I superconductor, this fullfils [25]

$$C_s - C_n = \frac{T_c}{\mu_0}\left(\frac{d}{dT}H_c(T)\right)^2_{T=T_c}, \quad (3)$$

where $\Delta C = C_s - C_n$ can be determined from the linear coefficient of the specific heat $\gamma_n$ from BCS theory. Figure 7 shows the relation in equation (3) for 22 elemental type I superconductors, plus a few select compounds with low critical fields (see supplementary material). The graph shows that equation (3) is well observed by known type I superconductors. The data point for $Sn_4As_3$ (shown in red) lies very close to the expected behaviour for a type I BCS superconductor in the weak coupling limit (solid black line). This leads us to conclude that $Sn_4As_3$ is a type I superconductor.

The low upper critical field on its own already provides strong indication that any triplet component of the order parameter is negligible in this system. Taken together with the observations of rather conventional SC in other non-centrosymmetric materials [4, 26], this does confirm that to observe a sizeable triplet component of the superconducting order parameter requires a material system where pairing is mediated by a mechanism other than electron–phonon coupling [27, 28].

## 5. Conclusions

We successfully synthesized $Sn_4As_3$ in the non-centrosymmetric crystal structure ($R3m$). Comprehensive characterisation of the normal state electronic structure shows metallic behaviour and only negligible influence of correlation effects. Our specific heat measurements are consistent with the weak-coupling BCS theory, with an SC gap of $\Delta = 0.177 \pm 0.003$ meV. This is in excellent agreement with the STM/STS measurements, which show a fully formed SC gap of $\Delta = 0.182 \pm 0.018$ meV, consistent with the material being a conventional superconductor with singlet pairing. Comparison of the specific heat and critical field shows that the material is a type I superconductor. In this material, the non-centrosymmetric crystal structure does not lead to unconventional superconducting properties and our findings show that spin–orbit coupling only plays a negligible role in its electronic structure.

## Acknowledgments

CAM and MJN acknowledge studentship funding from EPSRC under Grant no. EP/L015110/1. CMY and PW acknowledge funding through EP/S005005/1. CH acknowledges support by the Austrian Science Fund (FWF) Project No. P32144-N36 and the VSC-3 of the Vienna University of Technology. KSP, VAV, VMP, and AVM acknowledge support of the state assignment of the Ministry of Science and Higher Education of the Russian Federation (Project No. 0023-2019-0005). Work was done using equipment from the LPI Shared Facility Center. We thank Diamond Light Source synchrotron for access to beamline I05-ARPES (proposal number NT15663) that contributed to the results presented here. The research data underpinning this publication can be accessed at https://doi.org/10.17630/7bee9702-9d6b-4f05-9ca6-88b63465b511.

## ORCID iDs

C A Marques https://orcid.org/0000-0002-3804-096X
M D Watson https://orcid.org/0000-0002-0737-2814
L C Rhodes https://orcid.org/0000-0003-2468-4059
C Heil https://orcid.org/0000-0001-9693-9183
K S Pervakov https://orcid.org/0000-0003-1116-0897
V A Vlasenko https://orcid.org/0000-0002-4269-275X
V M Pudalov https://orcid.org/0000-0002-1992-193X
T K Kim https://orcid.org/0000-0003-4201-4462
P Wahl https://orcid.org/0000-0002-8635-1519